\documentclass[sigconf]{acmart}

\usepackage{footnote}
\usepackage{caption}
\usepackage[american]{babel}
\usepackage{csquotes}
\usepackage{array} 
\usepackage{booktabs}
\usepackage{siunitx}
\usepackage{amsmath}
\usepackage{lipsum}
\usepackage{dsfont}
\usepackage{comment}

\AtBeginDocument{%
  \providecommand\BibTeX{{%
    \normalfont B\kern-0.5em{\scshape i\kern-0.25em b}\kern-0.8em\TeX}}}

\def\BibTeX{{\rm B\kern-.05em{\sc i\kern-.025em b}\kern-.08emT\kern-.1667em\lower.7ex\hbox{E}\kern-.125emX}}
    
\acmDOI{}
\acmISBN{}
\acmArticle{}
\acmPrice{}

\acmConference[recsysXfashion'19]
{Workshop on Recommender Systems in Fashion, 13th ACM Conference on Recommender Systems}
{September 20, 2019}
{Copenhagen, Denmark}
\acmYear{2019}
\copyrightyear{2019}

\begin{document}

%
\title[Fashion Evaluation]{Assessing Fashion Recommendations: A Multifaceted Offline Evaluation Approach}

%

\author{Jake Sherman}
\affiliation{%
  \institution{True Fit}
  \city{Boston}
  \state{MA}}
\email{jsherman@truefit.com}

\author{Chinmay Shukla}
\affiliation{%
  \institution{True Fit}
  \city{Boston}
  \state{MA}}
\email{cshukla@truefit.com }

\author{Rhonda Textor}
\affiliation{%
  \institution{True Fit}
  \city{Boston}
  \state{MA}}
\email{rtextor@truefit.com}

\author{Su Zhang}
\affiliation{%
  \institution{True Fit}
  \city{Boston}
  \state{MA}}
\email{szhang@truefit.com}

\author{Amy A. Winecoff}
\affiliation{%
  \institution{True Fit}
  \city{Boston}
  \state{MA}}
\email{awinecoff@truefit.com}

%
\renewcommand{\shortauthors}{Sherman, et al.}

%
\begin{abstract}
Fashion is a unique domain for developing recommender systems (RS). Personalization is critical to fashion users. As a result, highly accurate recommendations are not sufficient unless they are also specific to users. Moreover, fashion data is characterized by a large majority of new users, so a recommendation strategy that performs well only for users with prior interaction history is a poor fit to the fashion problem. Critical to addressing these issues in fashion recommendation is an evaluation strategy that: 1) includes multiple metrics that are relevant to fashion, and 2) is performed within segments of users with different interaction histories. Here, we present our multifaceted offline strategy for evaluating fashion RS. Using our proposed evaluation methodology, we compare the performance of three different algorithms, a most popular (MP) items strategy, a collaborative filtering (CF) strategy, and a content-based (CB) strategy. We demonstrate that only by considering the performance of these algorithms across multiple metrics and user segments can we determine the extent to which each algorithm is likely to fulfill fashion users\textquotesingle{} needs.
\end{abstract}

%
%


%

%
\maketitle
\section{Introduction}

Few industries touch the lives and identities of consumers as intimately as fashion. Everyone makes decisions about what to wear, and these decisions reflect not only the prevailing cultural norms \cite{souiden2011cross}, but also the individual identity of the wearer \cite{casidy2009predicting, mulyanegara2009big}. Clothing can affect how the wearer feels and behaves as well as how others feel and behave in response to the wearer \cite{Johnson2014}. The motivations that drive fashion consumers include fashionability, individualization, self assurance, flaw minimization, and comfort, yet the importance of these motivations to fashion purchasing depends on the characteristics of the consumer \cite{Tiggemann2009}. That is, consumers\textquotesingle{} motivations for buying clothing are personal \cite{Vaccaro2016}. Consequently, any recommender system (RS) developed for the fashion domain should be evaluated on criteria that are specifically relevant to the needs of fashion consumers.

As with other domains, fashion users should receive ``accurate\textquotedblright{} recommendations (i.e., recommendations that are relevant), but accuracy alone is not sufficient. Although some researchers have been urging RS developers to think beyond accuracy for more than a decade (e.g., \cite{Ziegler2005}), accuracy is still the predominant focus of most RS evaluations. A survey of recent papers at the ACM RecSys Conference noted that while roughly 85\% of papers used some form of offline accuracy metric, a mere 20\% included a measure of diversity, novelty, or another alternative metric \cite{Jannach2016}. A dogged focus on maximizing accuracy can unintentionally degrade the end user experience. McNee, Riedl, and Konstan \cite{McNee2006} provide an illustrative example: a travel RS that recommends solely locations that users have previously visited would perform better on most accuracy metrics than a system that recommends novel travel destinations that are more interesting to the user. Thus, non-accuracy based evaluation measures are necessary for properly evaluating RS in general, and fashion RS specifically. 

Personalization is critical for good fashion recommendations. In marketing, ``personalization\textquotedblright{} is operationalized as the customization of goods and/or services to meet the needs of specific consumers \cite{Goldsmith1999}. Different fashion consumers have different motivations for purchasing fashion items \cite{Tiggemann2009}. As a result, one way to evaluate the personalization of fashion recommendations is by measuring how much recommendation lists vary from user to user (e.g., the approach in \cite{zhou2010solving}). In addition to list diversity, understanding the popularity bias in recommendations is important for evaluating personalization because recommendations dominated by popular items are necessarily depersonalized.   

We also want to understand how well RS perform within multiple user segments. Providing accurate and engaging recommendations is easier for users with rich interaction histories. However, within our own fashion datasets, few users have prior sales or even item views. That is, most users are new. Because personalization is so critical to fashion, approaches such as collaborative filtering (CF) that cannot provide recommendations for new users are unlikely to provide a good experience to most fashion shoppers. Consequently, we must assess how fashion RS perform with new as well as established users.  

The goal of our current work is to develop a methodology for more comprehensively evaluating the extent to which fashion RS produce quality recommendations. To provide an understanding of how different types of RS perform, we evaluate three algorithms: 1) a most popular (MP) items strategy, 2) a CF-based strategy, and 3) a content-based (CB) strategy. Our evaluation method consists of multiple measures of recommendation quality performed on multiple user segments.

\section{Related Works}
Several approaches have been developed to address the unique demands of providing fashion recommendations. Despite the prominence of the cold-start problem inherent in fashion data, some fashion recommendation approaches have nevertheless relied on CF. Hwangbo and colleagues \cite{Hwangbo2018} developed a novel user-based CF approach for recommending complementary and substitute fashion items that offers interesting algorithmic ideas, but is limited in that 1) only existing products are accommodated, and 2) recommendations are not personalized to users. Rue La La, a flash sale fashion retailer, developed a latent factors CF approach for providing fashion recommendations that overcomes the cold-start problem for items by recommending product groups to users instead of individual products \cite{Harrison2017}. Although this approach \textit{does} address the cold start problem for items, it \textit{does not} allow them to make personalized recommendations for new users.

Other methodologies for providing fashion recommendations eschew CF in favor of models that circumvent the cold start problem by leveraging user and/or product attributes to make recommendations. De Melo, Nogueira, and Guliato \cite{Melo2015} developed a content based (CB) fashion RS that constructs detailed clothing item attributes to build content profiles for each user and then uses k-nearest neighbors to make recommendations for new items. Although this approach overcomes the item cold-start problem, it cannot provide personalized recommendations for new users. A RS developed by Zalando \cite{Freno2017} leverages both user and product attributes within a learning to rank (L2R) framework, allowing them to make recommendations for both new items and users. However, Zalando only measured the accuracy of recommendations and did not evaluate how the system performed within different user segments. 

\section{Approach}

\begin{table*}[ht]
  \caption{Descriptive Statistics for Training Data}
  \label{tab:train}
\begin{tabular}{lllllll} \toprule

& Users & Products & Sales (\%) & Views (\%) & Unobserved (\%) \\ 
\midrule
\midrule

Retailer 1&  39,307 & 376 & 7,461(0.05\%) & 103,829(0.7\%) & 14,668,142(99.2\%)  \\ 
Retailer 2&  42,490 & 865 & 8,276(0.02\%) & 143,781(0.4\%) & 36,601,793(99.6\%)  \\
Retailer 3& 60,333 & 386 & 21,904(0.1\%) & 141,320(0.6\%) & 23,125,314(99.3\%)  \\ 

\bottomrule
\end{tabular}
\end{table*}

\begin{table*}[ht]
  \caption{Descriptive Statistics for Test Data}
  \label{tab:test}
\begin{tabular}{lllllllll} \toprule

& New Users (\%)& View Users (\%)& Sale Users (\%)& Products& Sales (\%)& Views (\%)& Unobserved (\%) \\ 
\midrule
\midrule

Retailer 1&  2,477(73.8\%) & 667(19.9\%) & 213(6.3\%) & 319 & 1,727(0.2\%) & 8,850(0.8\%) & 1,060,306(99.0\%)  \\ 
Retailer 2 &  6,048(69.0\%) & 1,997(22.8\%) & 720(8.2\%) & 676 & 5,171(0.1\%) & 50,443(0.9\%) & 5,869,526(99.1\%)  \\
Retailer 3 &  5,164(71.2\%) & 1,513(20.9\%) & 578(7.9\%) & 314 & 2,753(0.1\%) & 19,578(0.9\%) & 2,255,739(99.0\%)  \\ 

\bottomrule
\end{tabular}
\end{table*}

\subsection{Evaluation}
\subsubsection{Data Selection}We performed separate evaluations on three different retailers. Within retailers, we trained models for women\textquotesingle{}s dresses. We constructed each dataset by taking all user-product interactions that occurred in a one year period and split our data into training and test sets by allocating the first eight months to the training data and the remaining four months to the test data. Descriptive statistics for the training and test data are in Table \ref{tab:train} and Table \ref{tab:test}, respectively. In both the training and test data for all retailers, the overwhelming majority of observations in the user-item matrix are unobserved (i.e., the users did not view or buy the item). Our three retailers are also similar in terms of the distribution of users across our three user segments (see Section \ref{segmentation}).

\subsubsection{User Segmentation}\label{segmentation}One of our goals was to understand how RS would perform in user segments with different product interaction histories. Many of our users have no prior interaction history. Therefore, we define ``new users\textquotedblright{} as users who have no sales or views in the training data. Some users have viewed items in the training data, but never made a purchase. We consider these users ``view users\textquotedblright{}  (i.e., users with views in the training data, but no sales. Lastly, a minority of users have a prior purchase falling within the training data. We consider these users ``sale users.\textquotedblright{} We note that these user distinctions are based on the training data only. We perform our evaluations within each of these user segments as well as across all user segments to gain insight into the recommendation experience for different types of users.  

\subsubsection{Accuracy}Because we are primarily interested in how well RS perform at ranking items, we focus our evaluation on top-\textit{n} performance \cite{Valcarce2018a}. To assess model accuracy, we use a modified version of normalized discounted cumulative gain (NDCG) at \textit{k}. $NDCG_{k}$ is a normalized version of the discounted cumulative gain ($DCG_{k}$) metric, which is computed for a particular user as: 

\begin{equation}
  DCG_{k} = \sum_{i=1}^{k} \frac{rel_{i}}{log_{2}(i+1)} 
\end{equation}

\noindent where $rel_{i}$ is the relevance label for the $i^{th}$ item recommended to a user. $NDCG_{k}$ normalizes the $DCG_{k}$ by dividing it by the ideal $DCG_{k}$, or the $DCG_{k}$ that would be achieved by a perfect ranking. One of the limitations of typical $NDCG_{k}$ implementations is that if predictions are tied, the $NDCG_{k}$ value can be non-deterministic since the gain for tied items will be based on arbitrary ordering. To mitigate this problem, we implement the tie-aware $NDCG_{k}$ approach proposed in \cite{mcsherry2008computing}. We set \textit{k} to 10 as users typically see about 10 recommendations, and micro-average each user's $NDCG_{k}$ value together to report an aggregated value. In addition to reporting raw $NDCG_{k}$ values, we also report the percentage change between our $NDCG_{k}$ values and the $NDCG_{k}$ value that would result from a random ranking (\%$\Delta_{r}$) of the items since $NDCG_{k}$ will vary based on the number of items in data.

\subsubsection{Personalization}

To date, there is little consensus as to how best to measure recommendation diversity and personalization directly. We leverage two indirect measures that speak to diversity and personalization: an inter-user average distinct recommendations at \textit{k} metric and also a relative popularity metric. 

In order to measure inter-user diversity, we use the $AD_k$ (average distinct at $k$) metric, which is defined as:

\begin{equation}
  AD_{k} = \frac{1}{\frac{1}{2}(U^2 - U)} \cdot \sum_{i=1}^{U} \sum_{j=i+1}^{U} AD_{k,i,j}
\end{equation}

\noindent where $U$ is the total number of users, and $AD_{k,i,j}$, or the distinctness between a single pair of users, is defined as:

\begin{equation}
  AD_{k,i,j} = | L_{k,i} \bigtriangleup L_{k,j} | = | (L_{k,i} - L_{k,j}) \cup (L_{k,j} - L_{k,i}) |
\end{equation}

\noindent where $L_{k,i}$ is the set of top-\textit{k} recommended items for user $i$, and $L_{k,j}$ is is the set of top-\textit{k} recommended items for user $j$. $AD_{k,i,j}$ measures the cardinality of the symmetric difference between two different users\textquotesingle{} top-\textit{k} recommendations. In the case where two users\textquotesingle{} top-\textit{k} recommendations are exactly the same, the value of $AD_{k,i,j}$  will be zero. When they have no items in common, the value will be $2k$. In order to avoid the $O(U^2)$ complexity associated with computing $AD_{k}$ across the entire population of user pairs, we randomly sample the proportion$\frac{2}{U-1}$ of user pairs from the population of $\frac{1}{2}(U^2 - U)$ user pairs in order to create a randomly sampled set of user pairs. Then, we redefine $AD_k$ as:

\begin{equation}
  AD_{k} = \frac{2}{U-1} \cdot \sum_{i=1}^{U} \sum_{j=i+1}^{U} AD_{k,i,j} \cdot I_{i,j}
\end{equation}

\noindent where $I_{i,j}$ is an indicator variable that takes a value of 1 when the pair of users $(i,j)$ is in the randomly sampled set of user pairs, and a value of 0 otherwise.

In order to measure relative popularity, we use the $RP_k$ (relative popularity at \textit{k}) metric to quantify the popularity of users' top-\textit{k} recommendations relative to recommending the most popular $k$ items. $RP_k$ is defined as:

\begin{equation}
  RP_{k} = \frac{1}{U} \cdot \sum_{u=1}^{U} RP_{k,u}
\end{equation}

\noindent where $RP_{k,u}$, or the relative popularity at \textit{k} for a single user, is defined as:

\begin{equation}
  RP_{k,u} = \frac{\sum_{i=1}^{k} Q_{u,i}}{\sum_{i=1}^{k} Q_i}
\end{equation}

\noindent where $Q_{u,i}$ is the quantity sold of the top-$i^{th}$ recommendation for user $u$, and $Q_i$ is the quantity sold of the $i^{th}$ most popular product across all users. In the scenario where the top-\textit{k} most popular products are being recommended to all users, $Q_{u,i}$ becomes $Q_i$, resulting in a $RP_k$ value of one, its upper bound.

\subsection{Recommendation Algorithms}

\subsubsection{MP Recommendations}By definition, items are popular if they have broad appeal across a wide swath of users. Therefore, we might expect that by recommending popular items, we can achieve high levels of accuracy \cite{Cremonesi2010a}.  We determine which items are most popular based on sales. Specifically, we sum the total number of units sold for each item within our sales data on a retailer by retailer basis, and then recommend the top-\textit{k} items with the highest quantity of units sold. This MP recommendation strategy serves as a baseline algorithm that gives depersonalized but broadly palatable recommendations.

\subsubsection{CF Recommendations} Because some fashion RS use CF, we also include a CF-based recommendation algorithm. For our fashion items, we do not have explicit ratings of user preferences for products (e.g., a star rating of 1 to 5). Instead we must rely on implicit proxies for user preferences, in our case, product sales and views. In contrast to traditional item-based (e.g., \cite{sarwar2001}) and user-based CF (e.g., \cite{Resnick1994}), alternating least squares (ALS) is a matrix factorization CF strategy developed specifically for implicit feedback datasets \cite{Hu2008}. ALS allows user preference to be separated from confidence in user preference, which is useful for implicit datasets since indirect measures of user preferences are inherently noisy. Because sales can be considered a stronger signal of user preference than views, we weight sales more heavily in our model. We treat views as a binary for whether or not an item was viewed since multiple views may or may not indicate increased user preference. With our trained ALS model, we make predictions for all user-item combinations, and recommend the top-\textit{k} items with the highest predicted values.

\subsubsection{CB Recommendations}Many fashion RS use product and/or user attributes to give recommendations, so we also include a CB approach that leverages information about users and products. This CB recommendation strategy consists of two major phases. In the first phase, we fit an ALS CF to user sales and views and make predictions for user preferences. We then use these predictions to augment our original user-item interaction data. Specifically, if the user-item interaction was observed (i.e., was either viewed or sold), we retain the value of the original user-item interaction. If the user-item interaction was not observed, we substitute the value predicted by ALS. We then train a random forest model using the augmented outcomes as labels and information about users and products as features. For product features, we represent fashion details such as style attributes (e.g., dress shape, sleeve length) and price. For user features, we use fashion-relevant information about users such as body mass index (BMI), user age, and brand preferences. We train models separately for different retailers since the relationships between user and product features can be assumed to vary by retailer. Using the trained RF model, we recommend the top-\textit{k} items with the highest predicted values. 

\section{Results}

\begin{figure}[h]
  \centering
  \includegraphics[width=\linewidth]{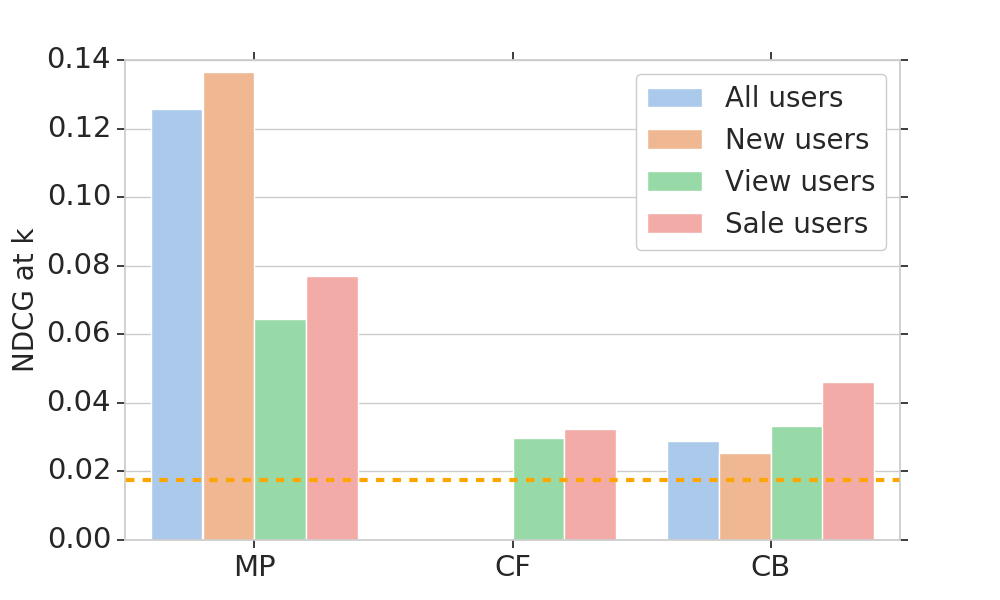}
  \caption{$NDCG_{10}$ for Retailer 1. The yellow dotted line corresponds to the $NDCG_{10}$ value for Retailer 1 that would result from a random ranking of the items.}
  \label{fig:ndcg}
  \Description{NDCG Figure}
\end{figure}

\begin{figure}[h]
  \centering
  \includegraphics[width=\linewidth]{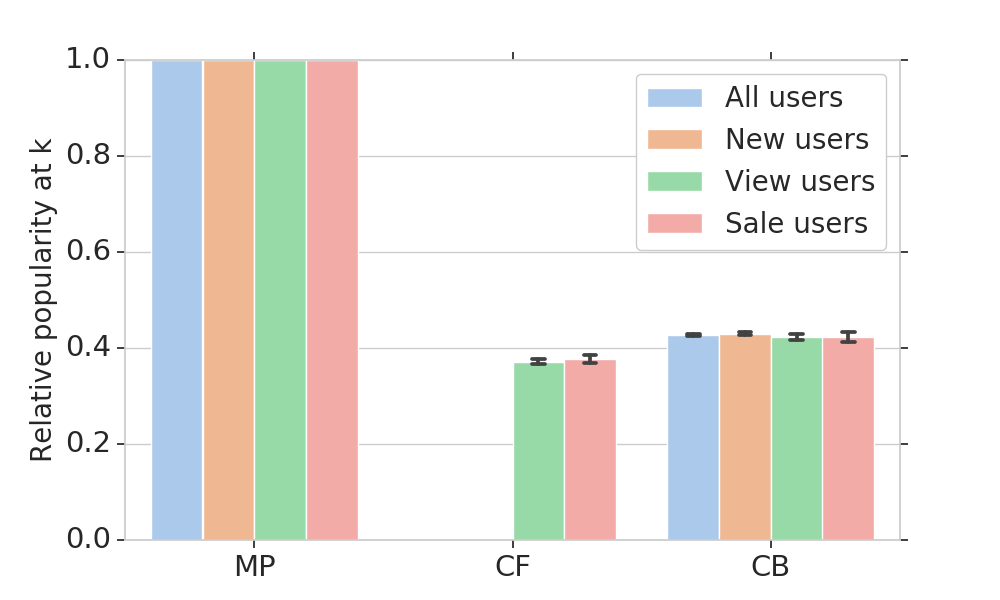}
  \caption{$RP_{10}$ for Retailer 1. Each error bar represents the 95\% confidence interval of the distribution of 1,000 bootstrap samples of $RP_{k,u}$ values. By only recommending the most popular items, the MP recommendation strategy always produces values of 1.}
  \label{fig:popularity}
  \Description{Popularity Figure}
\end{figure}

Results across retailers are presented in Tables \ref{tab:ndcg}-\ref{tab:pop}. To help illustrate the evaluation metrics and user segmentation, we provide depictions of results for Retailer 1 in Figures \ref{fig:ndcg}-\ref{fig:popularity}.

\subsection{Accuracy}

\begin{table}
  \caption{$NDCG_{10}$ Evaluation}
  \label{tab:ndcg}
\begin{tabular}{llll} \toprule

 & Retailer 1(\%$\Delta_{r}$)& Retailer 2(\%$\Delta_{r}$) & Retailer 3(\%$\Delta_{r}$) \\ 
\midrule
\midrule

Sale Users  & & &  \\ \midrule
 MP & 0.077(340.2\%) & 0.025(253.8\%) & 0.122(703.4\%) \\ 
 CF &  0.032(85.4\%) & 0.023(222.7\%) & 0.094(517.0\%) \\ 
 CB & 0.046(164.4\%) & 0.021(194.1\%) & 0.108(608.5\%) \\ 
\midrule  

View Users  & & &  \\ \midrule
 MP & 0.065(269.6\%) & 0.031(332.6\%) & 0.110(623.1\%) \\ 
 CF &  0.030(69.9\%) & 0.024(236.7\%) & 0.078(415.5\%) \\ 
 CB & 0.033(90.2\%) & 0.025(258.6\%) & 0.115(657.2\%) \\ 
\midrule  

New Users  & & &  \\ \midrule
 MP & 0.136(681.6\%) & 0.025(259.6\%) & 0.094(519.1\%) \\ 
 CF &  - & - & - \\ 
 CB & 0.025(46.0\%) & 0.012(68.6\%) & 0.089(485.8\%) \\ 
\midrule  

Average  & & &  \\ \midrule
MP & 0.126(620.3\%) & 0.026(259.8\%) & 0.102(569.2\%) \\ 
CF &  - & - & - \\ 
CB & 0.029(65.8\%) & 0.014(95.5\%) & 0.094(521.1\%) \\ 

\bottomrule
\multicolumn{4}{l} {\footnotesize Note: CF cannot make predictions for new users.}\\
\multicolumn{4}{l} {\footnotesize \%$\Delta_{r}$ is \% improvement over random.}\\

\end{tabular}
\end{table}

\noindent For $NDCG_{10}$, the MP recommendation strategy outperformed the CF and CB strategies. In general, CB recommendations outperformed CF recommendations on $NDCG_{10}$, with sale users in Retailer 2 being the only exception. Algorithm performance between user segments was dependent on retailer. For example, across all retailers, for CB recommendations, $NDCG_{10}$ was lowest for new users, but in Retailer 1, $NDCG_{10}$ for MP recommendations was higher for new than both view and sale users.  Overall, although MP was generally more accurate than CB, and CB was generally more accurate than CF, a finer grained analysis by user type and retailer revealed a more complex pattern of results. 

\subsection{Personalization}

\begin{figure}[h]
  \centering
  \includegraphics[width=\linewidth]{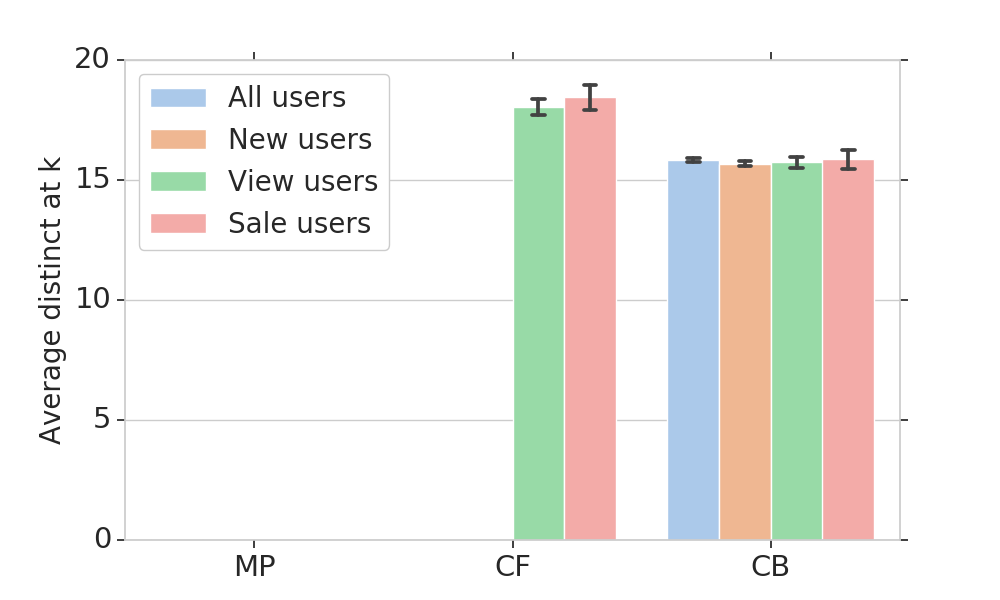}
  \caption{$AD_{10}$ for Retailer 1. Each error bar represents the 95\% confidence interval of the distribution of 1,000 bootstrap samples of $AD_{k,i,j}$ values. The MP recommendation strategy produces the same recommendations for all users, resulting in values of 0.}
  \label{fig:avgdist}
  \Description{Average distinct Figure}
\end{figure}

\begin{table}[htbp]
\setlength{\tabcolsep}{3pt}
  \caption{$AD_{10}$ Evaluation}
  \label{tab:avg}
\begin{tabular}{llll} \toprule

 & Retailer 1(SD) & Retailer 2(SD) & Retailer 3(SD) \\
\midrule
\midrule

Sale Users  & & &  \\ \midrule
 MP & 0(0) & 0(0) & 0(0) \\ 
 CF & 18.5(3.7) & 18.3(3.9) & 16.7(4.1) \\ 
 CB & 15.9(2.8) & 18.0(2.3) & 12.9(3.4) \\ 
\midrule  

View Users  & & &  \\ \midrule
 MP & 0(0) & 0(0) & 0(0) \\ 
 CF & 18.0(4.3) & 18.2(4.1) & 16.5(4.2) \\ 
 CB & 15.7(2.9) & 18.0(2.1) & 12.6(3.5) \\ 
\midrule  

New Users  & & &  \\ \midrule
 MP & 0(0) & 0(0) & 0(0) \\ 
 CF &  - & - & - \\ 
 CB & 	15.7(2.8) & 18.1(2.1) & 12.5(3.4) \\ 
\midrule  

Average  & & &  \\ \midrule
 MP & 0(0) & 0(0) & 0(0) \\ 
 CF &  - & - & - \\ 
 CB & 15.8(2.8) & 18.1(2.1) & 12.4(3.4) \\ 

\bottomrule
\multicolumn{4}{l} {\footnotesize SD is the standard deviation between user pairs.}\\

\end{tabular}
\end{table}

\noindent For $AD_{10}$, CF provided more distinctive recommendations than CB across all three retailers for the view and sale user segments. Meanwhile, MP provided the same popular recommendations to all users, resulting in $AD_{10}$ values of 0. Within each retailer/model combination, $AD_{10}$ exhibited very little variance across user segments. Figure~\ref{fig:avgdist} shows the increased distinctiveness of CF over CB and the low within-retailer/model variance in $AD_{10}$ across user segments for Retailer 1.

\begin{table}[ht]
\setlength{\tabcolsep}{3pt}
  \caption{$RP_{10}$ Evaluation}
  \label{tab:pop}
\begin{tabular}{llll} \toprule

 & Retailer 1(SD) & Retailer 2(SD) & Retailer 3(SD) \\ 
\midrule
\midrule

Sale Users  & & &  \\ \midrule
 MP & 1(0) & 1(0) & 1(0) \\ 
 CF & 0.38(0.06) & 0.31(0.07) & 0.43(0.13) \\ 
 CB & 0.42(0.08) & 0.24(0.06) & 0.55(0.13) \\ 
\midrule  

View Users  & & &  \\ \midrule
 MP & 1(0) & 1(0) & 1(0) \\ 
 CF & 0.37(0.06) & 0.31(0.07) & 0.43(0.12) \\ 
 CB & 0.42(0.08) & 0.24(0.06) & 0.56(0.12) \\ 
\midrule  

New Users  & & &  \\ \midrule
 MP & 1(0) & 1(0) & 1(0) \\ 
 CF &  - & - & - \\ 
 CB & 0.43(0.08) & 0.24(0.06) & 0.57(0.11) \\ 
\midrule  

Average  & & &  \\ \midrule
 MP & 1(0) & 1(0) & 1(0) \\ 
 CF &  - & - & - \\ 
 CB & 0.43(0.08) & 0.24(0.06) & 0.57(0.11) \\ 

\bottomrule
\multicolumn{4}{l} {\footnotesize Note: CF cannot make predictions for new users.}\\
\multicolumn{4}{l} {\footnotesize SD is the standard deviation across users.}\\

\end{tabular}
\end{table}

The MP recommendation strategy provided the most popularity-biased recommendations because by recommending the same, most-popular items to all users, MP always results in $RP_{10}$ values of 1. CB had more popularity-biased recommendations than CF for Retailers 1 and 3, while the opposite was true for Retailer 2. Overall, recommendations for Retailer 3 were the most popularity-biased, followed by Retailer 1, and then Retailer 2. Within each retailer/model combination, $RP_{10}$ exhibited very little variance across the four user segments.

\subsection{Summary of model results}

CB modeling was generally more accurate but less personalized than CF. Although CF generally provided more personalized recommendations for view and sale users compared with CB modeling, CF had much lower user-space coverage than CB modeling because CF cannot make recommendations for new users. While the MP recommendation strategy was able to provide very accurate recommendations, those recommendations were completely depersonalized, with the lowest possible $AD_{10}$ and highest possible $RP_{10}$. Only by performing a holistic evaluation that includes measures of personalization and evaluations across user segments are we able to expose the shortcomings of the MP and CF recommendation strategies. Despite having lower accuracy compared with MP and lower personalization when compared with CF, CB models were able to successfully balance accuracy with personalization while making recommendations to new users. 

\subsection{Summary of retailer results}

\begin{figure}[h]
  \centering
  \includegraphics[width=\linewidth]{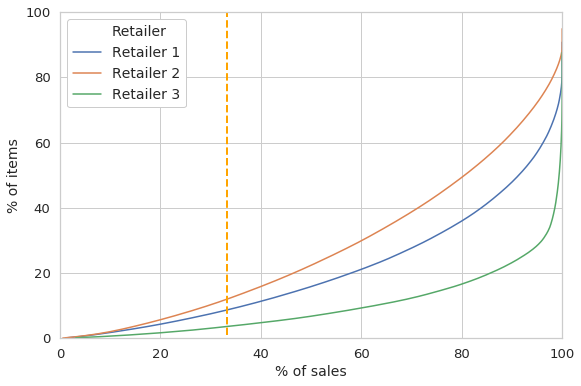}
  \caption{Sales distributions for our three retailers. Items are ordered by popularity, with the most popular items at the bottom. The set of popular items that make up a third of sales is known as the short-head, while the set of remaining items make up the long-tail \cite{Cremonesi2010a}. The yellow dashed line provides the demarcation between the items in the short-head and long-tail.}
  \label{fig:retailerpop}
  \Description{Average distinct Figure}
\end{figure}

\noindent Given the results by model type, we suspect that the patterns of results by retailer may be driven by differences in retailer sales distributions. In general, recommendations for Retailer 3 had the highest accuracy as measured by $NDCG_{10}$, but the lowest diversity and the highest popularity bias, which may be explained by Retailer 3 having the sales distribution most dominated by popular items. As shown in Figure \ref{fig:retailerpop}, one third of sales for Retailer 3 involve only the 3.7\% of most popular items, compared with Retailers 1 and 2, where one third of sales involve the 8.7\% and 12.1\% of most popular items, respectively. Additionally, in most cases CB modeling was more accurate but less personalized than CF, with the exception of higher $RP_{10}$ for CF than CB at Retailer 2, and lower $NDCG_{10}$ for CB than CF for sale users at Retailer 2. This exception may be explained by Retailer 2 having the sales distribution least dominated by popular items, where the accuracy and popularity bias of CB might be directly affected by the sales distribution of the underlying retailer data.

\section{Discussion}
Our goal was to propose an offline methodology for evaluating fashion RS. Because personalization is a critical feature of fashion, our evaluation framework includes accuracy as well as recommendation diversity and popularity bias. Moreover, because most users in our fashion datasets are new, we performed our analyses separately for users based on prior interaction history. By considering multiple metrics within multiple user segments, we gain a better understanding of how algorithm decisions are likely to influence the experience of the end users. 

Although our results varied to some extent by user segment and retailer, we can still make several important conclusions. First, across all of our retailers, our data is very sparse. For comparison, the Netflix dataset and the MovieLens dataset, both of which have been used extensively for RS research \cite{Bennett2007, F.Maxwell2015}, demonstrate denser data than any of our three retailers. The overwhelming majority of users represented in the test dataset had no views or sales in the training dataset. As a result, our CF algorithm was unable to provide recommendations for over 70\% of users, making CF a poor algorithm choice for fashion. In contrast, our MP algorithm was able to provide accurate recommendations for all user segments; however, because item popularity was calculated across all users, the MP algorithm provides no personalization. Our CB approach represents the best algorithm choice of the three because it provides: 1) relatively accurate recommendations, 2) an acceptable level of personalization, and 3) complete user-space coverage. 

Our fashion RS evaluation approach has many advantages over more simplistic approaches; however, there are several ways in which our approach is limited. Here, we define users based on interaction history, but user groups could be defined along many axes (e.g., demographics, frequent versus infrequent shoppers). Also, our approach focused on segmenting users, not products. Content providers may also be interested in how well RS perform within specific subsets of their products (e.g., new versus classic products). Furthermore, we limited our RS comparisons to three relatively basic algorithms. Comparing different variants of these algorithms (e.g., neighborhood-based CF versus model-based CF) could provide additional nuance to our results. Future research could apply our evaluation approach to more variants of common algorithms as well as to novel algorithms specifically tailored to fashion recommendation (e.g, an algorithm focused on flaw minimization or comfort).

Here, we have proposed a more comprehensive offline evaluation. However, prior research has indicated that offline and online metrics are not always correlated \cite{beel2013comparative}, calling into question the utility of offline evaluation. One of reasons why offline and online metrics disagree could be that most offline evaluation methods are singularly focused on accuracy \cite{beel2013comparative} and as a result, fail to capture the full range of human factors that influence users\textquotesingle{} experiences. A multifaceted evaluation approach applied to multiple user segments is more likely to promote algorithms that perform well on online metrics (e.g., click through rates, increased sales, etc.,). Nevertheless, an important next step will be performing an online evaluation to validate our offline results. 

In sum, our current work demonstrates the importance of evaluating recommendations from multiple angles. By performing a multifaceted offline evaluation, we can develop a better insight into how our RS are likely to perform when encountered by real-world fashion users.

\bibliographystyle{ACM-Reference-Format}
\bibliography{FashionXBib}

\end{document}